# The fast C($^3$P) + CH$_3$OH reaction as an efficient loss process for gas-phase interstellar methanol

*Robin J. Shannon,$^a$ Christophe Cossou,$^{b,c}$ Jean-Christophe Loison,$^{d,e}$ Philippe Caubet,$^{d,e}$ Nadia Balucani,$^f$ Paul W. Seakins,$^{a,g}$ Valentine Wakelam,$^{b,c}$ and Kevin M. Hickson$^{d,e}$\**

**Corresponding Author**

\* Correspondence to: km.hickson@ism.u-bordeaux1.fr

Rate constants for the C($^3$P) + CH$_3$OH reaction have been measured in a continuous supersonic flow reactor over the range 50 K ≤ $T$ ≤ 296 K. C($^3$P) was created by the *in-situ* pulsed laser photolysis of CBr$_4$, a multiphoton process which also produced some C($^1$D), allowing us to investigate simultaneously the low temperature kinetics of the C($^1$D) + CH$_3$OH reaction. C($^1$D) atoms were followed by an indirect chemiluminescent tracer method in the presence of excess CH$_3$OH. C($^3$P) atoms were detected by the same chemiluminescence technique and also by direct vacuum ultra-violet laser induced fluorescence (VUV LIF). Secondary measurements of product H($^2$S) atom formation have been undertaken allowing absolute H atom yields to be obtained by comparison with a suitable reference reaction. In parallel, statistical calculations have been performed based on *ab-initio* calculations of the complexes, adducts and transition states (TSs) relevant to the title reaction. By comparison with the experimental H atom yields, the preferred reaction pathways could be determined, placing important constraints on the statistical calculations. The experimental and theoretical work are in excellent agreement, predicting a negative temperature dependence of the rate constant increasing from 2.2 × 10$^{-11}$ cm$^3$ molecule$^{-1}$ s$^{-1}$ at 296 K to 20.0 × 10$^{-11}$ cm$^3$ molecule$^{-1}$ s$^{-1}$ at 50 K. CH$_3$ and HCO are found to be the major products under our experimental conditions. As this reaction is not considered in current astrochemical networks, its influence on interstellar methanol abundances is tested using a model of dense interstellar clouds.

1. Introduction

The reactions of ground state atomic carbon, $C(^3P)$, in the gas-phase have been the focus of extensive experimental kinetic[1-5] and dynamical[6-10] studies over a wide range of temperatures and energies. Such processes are important in a diverse range of environments ranging from combustion systems at high temperature to the chemistry of interstellar clouds at low temperature (10 K). Given its high reactivity (rate constants greater than $10^{-10}$ cm$^3$ molecule$^{-1}$ s$^{-1}$) towards unsaturated hydrocarbon molecules, numerous investigations of $C(^3P)$ reactions with these coreagents in particular have improved our understanding of the mechanisms involved in such barrierless insertion type reactions. In stark contrast, there are relatively few studies of the reactions of $C(^3P)$ with saturated coreagent species and no earlier experimental studies exist of the gas-phase reactions of aliphatic alcohols at any temperature. Such reactions are particularly relevant to alcohol combustion models in addition to modelling studies of interstellar environments.

Gas-phase methanol CH$_3$OH, is observed throughout the interstellar medium and is present under a wide range of physical conditions.[11-13] Methanol can be formed directly in the gas-phase through a mechanism initially involving a radiative association reaction followed by dissociative recombination

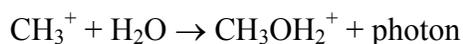
$CH_3^+ + H_2O \rightarrow CH_3OH_2^+ + \text{photon}$

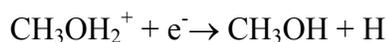
$CH_3OH_2^+ + e^- \rightarrow CH_3OH + H$

with a measured product yield of 3%.[14] Nevertheless, this mechanism is not efficient enough to explain observed gas-phase methanol abundances. Methanol is also widely detected in the form of methanol ice[15] at high abundances relative to water ice. Recent experiments have shown that the successive hydrogenation of adsorbed CO molecules[16,17] is an effective methanol formation mechanism. One or several efficient desorption mechanisms must then be invoked to return methanol to the gas-phase.[18] Observations of high gas-phase methanol abundances ($10^{-6}$ - $10^{-7}$ relative to total hydrogen nH + 2nH$_2$) in hot cores,[11] the dense regions of gas surrounding protostars, tend to support this hypothesis. In these regions, temperatures are generally in the range 20 - 300 K allowing methanol ice trapped on interstellar grains to sublimate. In contrast, in dense clouds where temperatures of 10 K prevail and the corresponding desorption mechanisms are less efficient, methanol abundances are typically much lower ($10^{-9}$), although they are generally reproduced by dense cloud models including gas-grain chemistry if reactive desorption mechanisms are employed.[19]

Observations of ground state atomic carbon by emission spectroscopy through its ground state fine structure line emission at 492 GHz[20,21] indicate that it could be one of the most abundant species in dense interstellar clouds so that current models[22] adopt peak gas-phase C($^3$P) abundances equal to $10^{-4}$ – $10^{-5}$. Given the low reactivity of CH$_3$OH towards other abundant interstellar species such as He, H, N, O, H$_2$ and CO, the reaction between C and CH$_3$OH could represent the major loss process for interstellar methanol were it found to occur rapidly at low temperature.

A recent quantum chemical investigation,[23] at the CCSD(T)/cc-pVTZ level of theory, of the pathways involved in the C($^3$P) + CH$_3$OH system, predicted that reaction occurs through the formation of one of two initial Van der Waals complexes CH$_3$O(C)H or (C)CH$_3$OH formed by addition of atomic carbon to either the oxygen atom or the carbon atom of CH$_3$OH respectively. The former complex was found to be lower in energy than the reagents (-47 kJ mol$^{-1}$) with the latter being above the reagent level (+6 kJ mol$^{-1}$). These complexes could then evolve through TSs which were all predicted to be higher in energy than the reagent level for all possible exit channels so that back dissociation of the complex was thought to be the major outcome at low temperatures. Nevertheless, two of the exit channels from the CH$_3$O(C)H complex are seen to have TS energies only slightly higher than the reagent level taking into account zero point energy; one leading directly to the formation of bimolecular products CH$_3$ + COH (TS energy +22 kJ mol$^{-1}$) with the other resulting in the formation of an adduct CH$_3$OCH (TS energy +5 kJ mol$^{-1}$). The exit channel from the (C)CH$_3$OH complex was predicted to lead to the CHCH$_2$OH adduct through a TS only 1 kJ mol$^{-1}$ higher than the reagent level.

Given the low energies of the exit TSs from the (C)CH$_3$OH and CH$_3$O(C)H Van der Waals complexes and the high uncertainty generally associated with such calculations, further experimental and theoretical investigation of this reaction is warranted. Here we present the results of a combined experimental and theoretical study of this reaction. Experimentally, a supersonic flow reactor using pulsed laser photolysis for C($^3$P) production coupled with both direct laser induced fluorescence and indirect chemiluminescence detection was used to follow the kinetics of C($^3$P) loss in the presence of an excess of CH$_3$OH over the 50 K – 296 K range. In addition, secondary measurements of the formation of H atom co-products were also performed to elucidate the most favoured reaction pathways. Theoretically, the experimental results are complemented by statistical calculations of the reaction using the MESMER (Master Equation Solver for Multi Energy Well Reactions) code to predict the low temperature rate constants and product branching ratios of the title reaction.[24-26] MESMER utilizes the energies and structures of transition states and complexes calculated by quantum chemical calculations specifically undertaken for the present investigation.

## 2. Methodology

**Experimental methods**

Measurements were performed using a miniaturized continuous supersonic flow reactor based on the original apparatus designed by Rowe *et al.*[27] As the cold supersonic flow generated by the Laval nozzle is isolated from the walls of the reaction vessel in such experiments, this method is ideally suited for the study of reactions involving reagents which have low saturated vapour pressures at low temperature. In particular, species such as methanol which are in the liquid phase at room temperature can be introduced into the cold supersonic flow at concentrations well above their saturation limit. A detailed description of the method has been provided elsewhere[28,29] so only experimental details specific to the present investigation will be outlined here. Five different Laval nozzles were employed in the present study giving access to six different temperatures and densities through the use of different carrier gases. Room temperature measurements were performed by removing the nozzle and by significantly reducing the flow velocity in the reactor to eliminate pressure gradients in the observation region, effectively using the apparatus as a conventional slow flow reactor. The temperature, density and velocity of the supersonic flows were calculated from separate measurements of the impact pressure and the stagnation pressure within the reservoir. The measured and calculated values are summarized in Table 1 alongside other relevant information.

C($^3$P) atoms were generated by the multiphoton dissociation of $CBr_4$ molecules using unfocused 266 nm radiation with ~ 23 mJ of pulse energy. The photolysis laser was aligned along the supersonic flow, to create a column of carbon atoms of uniform density. $CBr_4$ was entrained in the flow by passing a small amount of carrier gas over solid $CBr_4$ held at a known pressure. An upper limit of $2 \times 10^{13}$ molecule $cm^{-3}$ was estimated for the gas-phase concentration of $CBr_4$ in the experiments from its saturated vapour pressure at room temperature.

Methanol was introduced into the flow upstream of the Laval nozzle using a controlled evaporation mixing (CEM) system. A 1 litre reservoir containing methanol was held at a pressure of 2 bar relative to atmospheric pressure. It was connected to one of two different liquid flow meters to allow flows of between 0.1 and 5 g $hr^{-1}$ or between 0.02 and 1 g $hr^{-1}$ of liquid methanol to be passed into an evaporation device heated to 353 K. A fraction of the main carrier gas flow (Ar or $N_2$) was introduced into the CEM through another mass flow controller, carrying methanol vapour into the reactor. To determine the gas-phase methanol concentration, the CEM output passed through a 10 cm absorption cell held at room temperature. The 185 nm line of a mercury pen-ray lamp was used to determine the gas-phase methanol absorption in the cell. The transmitted intensity was measured alternately in the presence and absence of methanol vapour to yield values of the attenuated and non-attenuated intensities, I and $I_0$ respectively using a solar blind channel photomultiplier tube (CPM) operating in photon counting mode with a spectral response in the 115 - 200 nm range. Typically, the number of photons falling on the CPM per second was recorded at one second intervals throughout the entire duration of the kinetic measurement. The same procedure was used in the absence of methanol to obtain $I_0$ both before and after the kinetic measurement to assess the magnitude of any potential variations of the lamp intensity. Unfortunately, it was also observed that the detector was slightly sensitive to short wavelength UV photons, although methanol itself has a negligible absorption cross-section at wavelengths greater than 220 nm. As the 254 nm mercury line is very much more intense than the 185 nm one, the measured values of I and $I_0$ therefore contained a significant constant contribution from 254 nm photons. Consequently, calibration measurements were performed on a daily basis to evaluate the 254 nm contribution to the main experiments. During these calibrations a fixed methanol vapour flow was introduced into the absorption cell and values of I were recorded over a range of pressures ($I_0$ remained constant). The gas-phase methanol concentrations calculated from the measured I and $I_0$ values using the absorption cross-section of methanol vapour at 185 nm ($6.05 \times 10^{-19}$ cm$^2$)[30] and the Beer-Lambert law were then plotted against the total pressure and different trial values for the constant were subtracted from both I and $I_0$ values until the methanol concentration versus total pressure plot yielded a straight line that passed through the origin. This correction factor, which represented approximately 47% of the total unattenuated intensity was then subtracted from all of the I and $I_0$ values recorded during the experiments.

The output of the cell was connected to the CRESU reactor using a heating hose maintained at 353 K to avoid condensation. As the methanol vapour was diluted by at least a factor of five on entering the room temperature nozzle reservoir through mixing with the main carrier gas flow, we assume that no supplementary condensation losses occurred upstream of the Laval nozzle. In this way, gas-phase methanol concentrations as high as $6.6 \times 10^{14}$ molecule cm$^{-3}$ could be obtained in the cold flow.

The temporal profile of atomic carbon was followed using two different methods. In earlier experiments, C($^3$P) was detected indirectly by chemiluminescent emission of NO through its highly exothermic reaction with NO$_2$ added to the flow

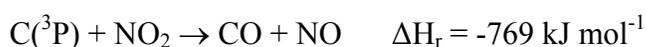

$$C(^3P) + NO_2 \rightarrow CO + NO \quad \Delta H_r = -769 \text{ kJ mol}^{-1}$$

The NO$_2$ concentration was constant for any single series of measurements with a maximum estimated value of $5 \times 10^{13}$ molecule cm$^{-3}$. As the NO molecules (denoted NO*) are produced over a range of electronic states with transitions corresponding to the γ, β, δ and ε bands from 220-450 nm,[31] we chose to observe NO* emission occurring in the 300-400 nm range (corresponding mostly to transitions originating in the β B$^2\Pi_r$ - X$^2\Pi_r$ band) using a bandpass filter and a UV sensitive photomultiplier tube (PMT). NO* detection allowed us to trace the entire C($^3$P) temporal profile for each individual laser shot, drastically reducing the time required for signal acquisition. The time dependent chemiluminescence signals consisted of 1000 time points acquired by a 500 MHz digital oscilloscope. The signal was averaged over 768 photolysis laser cycles for each kinetic decay, the characteristic decay time for each trace depending on the methanol concentration in the supersonic flow. The measurements were then repeated for a range of excess methanol concentrations at different temperatures.

In later experiments, C($^3$P) was detected directly through on-resonance VUV LIF using the $2s^22p^2$ $^3P_2 \rightarrow 2s^22p3d$ $^3D_3^0$ transition at 127.755 nm and a solar blind PMT operating in pulsed mode coupled to a boxcar integration system through a fast preamplifer. For this purpose, a Nd:YAG pumped dye laser operating at 766.5 nm was frequency doubled in a type I beta barium borate crystal to produce UV radiation at 383.3 nm. The resulting UV beam was then steered and focused into a cell containing 30 Torr of xenon gas mounted at the level of the reactor observation axis, perpendicular to both the supersonic flow and the PMT. The tunable VUV radiation produced by third harmonic generation of the UV beam was collimated by an $MgF_2$ lens positioned at the exit of the cell before interacting with the cold carbon atoms in the supersonic flow.

Given the multiphotonic nature of the $CBr_4$ photodissociation process, it was also possible that carbon atoms in the first excited state, $C(^1D)$, could be produced ($C(^1D)$ is only 122 kJ mol$^{-1}$ higher than $C(^3P)$ with each 266 nm photon providing 450 kJ mol$^{-1}$ of energy) which might interfere with measurements of the ground state carbon atom reactions. In order to check for $C(^1D)$ production, two different methods were employed. Firstly, we attempted to detect $C(^1D)$ directly through its $2s^22p^2$ $^1D_2$ → $2s^22p7d$ $^1F_3^0$ transition at 127.498 nm using the VUV generation method described above. Although some fluorescence was observed over a longer timescale to the $C(^3P)$ one, the signal was too weak to be able to perform quantitative measurements (this transition has an Einstein A coefficient which is 10 times weaker than the one used to detect $C(^3P)$[32]). Secondly, as $C(^3P)$ atoms are unreactive with molecular hydrogen at low temperature, whereas $C(^1D)$ atoms react rapidly to produce CH and H,[33] we performed a series of test experiments whereby $H_2$ was added to the supersonic flow in the absence of other coreagent species to check for H atom formation. In these experiments, the frequency tripling setup described above was adapted, using a fundamental dye laser wavelength of 729.4 nm coupled with 45 Torr of krypton in the tripling cell to detect $H(^2S)$ atoms by VUV LIF via the Lyman α transition at 121.567 nm. H atom production was observed, confirming that the 266 nm photolysis of $CBr_4$ results in at least some $C(^1D)$ formation. The $C(^1D)/C(^3P)$ ratio was determined by comparing relative product H atom yields from the $C(^1D) + H_2$ and $C(^3P)$ + methylacetylene, $C_3H_4$, reactions. The

C($^3$P) + C$_3$H$_4$ reaction is known to produce atomic hydrogen at room temperature at the 0.85 ± 0.07 level.[34] In this case, we considered that the H atom branching ratio from the C($^3$P) + C$_3$H$_4$ reaction did not change at 127 K and that the corresponding C($^1$D) + C$_3$H$_4$ reaction also yielded 100% H atoms. This latter assumption was verified by test experiments in which identical H atom intensities from the C($^3$P)/C($^1$D) + C$_3$H$_4$ reactions were recorded in the presence and absence of a large excess H$_2$ concentration ([H$_2$] ≥ 3.2 × 10$^{14}$ molecule cm$^{-3}$), allowing us to check that all of the C($^1$D) was converted to atomic hydrogen in both cases. The C($^1$D)/C($^3$P) ratios were determined to be (0.15 ± 0.04) and (0.10 ± 0.02) at 296 K and 127 K respectively, both values representing weighted averages of five individual ratio measurements. Subsequent test measurements to check for potential interferences in the main experiments from C($^1$D) reactions or relaxation back to the ground state are outlined in the kinetic results section.

The H atom VUV LIF setup was also used in secondary experiments performed at 127 K and 296 K investigating H($^2$S) formation from the C($^3$P)/C($^1$D) + CH$_3$OH reactions. As before, test experiments of these reactions were performed in the presence and absence of excess H$_2$, allowing us to evaluate the H atom yield from the C($^1$D) + CH$_3$OH reaction to be approximately 100%. The H atom fluorescence signals from the C($^3$P)/C($^1$D) + CH$_3$OH and C($^3$P)/C($^1$D) + C$_3$H$_4$ reactions were thus compared to obtain the H atom yield from the C($^3$P) + CH$_3$OH reaction. The results of the H atom yield measurements which are presented in the product branching ratio section provided important constraints for the statistical calculations.

VUV LIF signals from the reagent C($^3$P) (or product H($^2$S)) atoms were recorded as a function of delay time between photolysis and probe lasers. For a given excess coreagent (H$_2$, CH$_3$OH or C$_3$H$_4$) concentration, 30 datapoints were accumulated at each time interval with temporal profiles consisting of a minimum of 49 time intervals. Several time points were also recorded by firing the probe laser before the photolysis laser to establish the pre-trigger baseline value. This procedure was repeated several times for each value of the coreagent concentration producing a minimum of 23 individual decay profiles for each temperature. Both the LIF and chemiluminescence intensities were measured at a fixed distance from the Laval nozzle; the chosen distance corresponding to the maximum displacement from the nozzle possible whilst still maintaining optimal flow conditions so that the decays could be exploited over as large a time period as possible.

All gases were flowed directly from cylinders with no further purification prior to usage (Linde: Ar 99.999%, $SF_6$ 99.999%, Kr 99.99%, Xe 99.95%; Air Liquide: $N_2$ 99.999%, $NO_2$/He 5%). The carrier gas and precursor flows were all passed into the reservoir via digital mass flow controllers. The controllers were calibrated using the pressure rise at constant volume method for the specific gas used.

**Quantum chemical calculations**

Previous calculations[23] of the $C(^3P)$ + $CH_3OH$ reaction predicted that all pathways should be characterized by barriers, the smallest one being equal to 5 kJ mol$^{-1}$, which was inconsistent with our preliminary low temperature experimental results. To get a better description new theoretical calculations were performed, including the Davidson corrected multi-reference configuration interaction (MRCI + Q) and multi-reference perturbation theory (CASPT2) with complete active space self-consistent field (CASSCF) wavefunctions, to take into account the eventual multi-configurational aspect of this reaction. In addition, DFT calculations were also performed at the M06-2X/cc-pVQZ level. This highly non-localized M06-2X functional[35] is well suited for predicting structures and energetics of the TSs. Finally, calculations were also performed at the CCSD(T)/aug-cc-pVTZ level and also using the explicitly correlated approximation CCSD(T)-F12. MRCI+Q, CASPT2 and CCSD(T) calculations were carried out with the MOLPRO 2010 package and DFT calculations were performed with the Gaussian09 package. The CASSCF, MRCI and CASPT2 calculations were performed with 12 electrons distributed in 10 orbitals with the 1s and 2s orbitals of carbon and oxygen kept doubly occupied but fully optimized. We verified that the energy change using different active spaces was smaller than the energy change when using different calculation methods. The CASSCF calculations lead to mono-configurational wavefunctions for adducts and TSs (with the exception of TS3, (for clarity we use the same notation as the one used by Dede & Ozkan),[23] then CCSD(T) calculations should be reliable as indicated by the T1 and D1 diagnostic values (T1 values are in the 0.009 - 0.012 range and D1 values are in the 0.020 - 0.040 range). The geometries for stationary points were fully optimized for each method and frequencies were calculated at the DFT and CCSD(T) levels. For each TS, characterized by one imaginary frequency (first order saddle points) on the PES, we determined the minimum energy pathways (MEPs) performing intrinsic reaction coordinate analyses (IRC) at the DFT level (M06-2X/cc-pVQZ). A schematic energy diagram of the various reaction pathways is shown in Figure 1 whilst the calculated stationary point energies are summarized in Table S1. The geometries and frequencies of the structures of complexes 1 and 2 and TSs 1, 3 and 4 are listed in the supporting information file.

The $C(^3P_{0,1,2})$ + $CH_3OH(^1A')$ reaction leads to 3 surfaces in the entrance valley, one $^3A'$ surface and two $^3A''$ surface when the C atom approaches in $C_s$ geometry or 3 $^3A'$ surfaces when the carbon atom approaches in $C_1$ geometry (no symmetry). All further calculations were performed with $C_1$ symmetry. At the CASPT2 level with CASSCF geometry optimized for non-relaxed $CH_3OH$, only the first surface is attractive, leading to two different van der Waals complexes. The first complex (complex1) where the carbon atom is attached to the oxygen atom is strongly bound with a C-O bond length between 1.69 Å and 2.11 Å depending on the method used. The second complex (complex2), found only at the DFT and CCSD(T) levels, is closer in energy to the reagent level, the carbon atom being bound to an hydrogen atom of the methanol $CH_3$ group with a C-H bond length close to 1.5 Å. It should be noted that complex2 could not be found using the CASSCF method. Further evolution of complex1 leads to a $^3CH_3OCH$ adduct through TS1 corresponding to C insertion into the O-H bond, or directly to $CH_3$ + HOC products through TS3. Complex2 leads to the $CHCH_2OH$ adduct through TS4 corresponding to C insertion into the C-H bond. We did not examine the potential for insertion into the C-O bond as previous calculations have found that these pathways are characterized by transition states at much higher energies.[23] It is worth noting that C insertion into the O-H and C-H bonds have similar TS energies, close to the entrance level at the CCSD(T)/avtz level. Moreover, the TS1 and TS4 energies are very different from the corresponding TSs for insertion of a carbon atom into the C-H bond for the C + $CH_4$ reaction (barrier calculated to be +51 kJ mol$^{-1}$ at the CCSD(T)/6-311G level[36]) and also for the O-H insertion of a carbon atom for the C + $H_2O$ reaction (barrier calculated to be +30 kJ mol$^{-1}$ at the CCSD(T)/vtz level.[37]

As shown in Table S1, the calculated complex and TS energies are very method dependent, for example varying by 81 kJ mol$^{-1}$ for TS4. Our results are in agreement with previous calculations[23] as well as with expectations on the accuracy of the methods. Indeed, as shown by Wagner,[38] DFT methods usually underestimate energy barriers (even if the M06-2X functional leads to results in better agreement with experimental data than B3LYP). The reaction enthalpy, $\Delta H_r$, for the $CH_3$ + HCO exit channel calculated at the M06-2X level is in very good agreement with thermochemical data[39] (-326 kJ mol$^{-1}$ in both cases). In contrast, MRCI methods usually overestimate energy barriers whereas CASPT2 calculations are likely to underestimate them (for example MCQDPT2 calculations lead to lower complex and TS energies than CCSD(T) methods for the C + $H_2O$ reaction[40]). In this case, RCCSD(T) and RCCSD(T)-F12 methods should lead to the more accurate values considering the T1 and D1 diagnostic values[41,42] in agreement with the fact that CASSCF calculations lead to mono-configurational wavefunctions. The calculated TS energies themselves are expected to have large associated uncertainties (while it is difficult to provide a precise estimate, the TS energies could be as much as 10 kJ mol$^{-1}$ higher or lower than the calculated value). The uncertainties on the well depth energies of complex1 and complex2 should be somewhat smaller than the TS one.

**Statistical calculations**

Rate constants and product branching ratios were calculated using the Master Equation solver MESMER,[24] an open source program which uses matrix techniques to solve the energy grained chemical master equation for a series of unimolecular systems connected by various transitions states and local minima leading from reagents to products. Briefly, the phase space for all wells on the potential energy surface was divided into grains of a set size and isoenergetic, microcanonical rate coefficients for reactive processes between different species were calculated using RRKM theory, whilst energy transfer probabilities between different grains of the same species were obtained using an exponential down model. The time resolved change in the grain populations could then be described using a set of coupled differential equations with the form given below:

$$\frac{d}{dt}\mathbf{p} = \mathbf{M}\mathbf{p}$$

where p is the population vector containing the populations, $n_i(E,t)$, of the energy grains, $i$ refers to the $i$th isomer, and M is the matrix that determines grain population evolution as a result of collisional energy transfer and reaction. Phenomenological rate coefficients for the system were then obtained through a modified version of the approach used by Bartis and Widom.[43,44]

For all species, external rotations were considered to be rigid rotors whilst most vibrational modes were treated as harmonic oscillators. However there were a number of large amplitude torsions in the current work which were treated as separable hindered rotations. For these motions, torsional potentials were obtained through fully relaxed scans about the corresponding dihedral angle using the M06-2X/6-31+G** level of theory. Using discrete points from these potentials, energy levels for the hindered rotations were calculated using MESMER as previously described.[24]

For the current calculations a grain size of 50 cm$^{-1}$ was used above 150 K though this was reduced to a value as low as 10 cm$^{-1}$ at the lowest temperatures. Harmonic frequencies and rotational constants were taken from the M06-2X/cc-pVQZ calculations but given the large range of energies for the stationary points between the different levels of theory used here, some of the energies were varied during the master equation calculation as will be discussed in the results and discussion section. In preliminary calculations it was found that treating the reactions via TS1, TS3 and TS4 as irreversible introduced less than 0.01% error into the calculations. As such, the post transition state adducts were treated as sinks for the majority of the calculations.

This system includes two bimolecular source terms corresponding to the associations of $C(^3P)$ and $CH_3OH$ to form complex1 and complex2. For these two barrierless reaction steps, an inverse Laplace transform (ILT) method[45] was used to convert high pressure limiting canonical rate constants for the association into microcanonical rate coefficients assuming the Arrhenius form:

$$k_a = A\left(\frac{T}{T_0}\right)^n e^{-E_a/k_b T}$$

In initial master equation calculations a temperature independent capture rate of $2.5 \times 10^{-10}$ cm$^3$ molecule$^{-1}$ s$^{-1}$ was assumed for both associations. However, as will be discussed, from these initial calculations combined with the experimental data, it became apparent that the product channel via complex1 and TS1 was the dominant one. In light of this, more detailed variational calculations were performed to give an *a priori* determination for the high pressure limiting capture rate to form complex1. These calculations were performed using the multifaceted variable reaction coordinate (VRC) methodology,[46,47] as implemented in Polyrate 2008.[48] Such calculations require a large amount of potential information to obtain sums of states for the transitional modes in the $C(^3P) + CH_3OH$ association through phase space integration and this information was obtained through "on the fly" calculations at the M06-L/6-31G* level of theory using the Gaussian09 software package interfaced with Polyrate using Gaussrate 2009-A.[49] In these calculations a total of three pivot points were used. One was centred upon $C(^3P)$ and the other two were centred upon the O and H atoms of the methanol hydroxyl group.

## 3. Results and Discussion

**Product branching ratios**

Typical temporal profiles of the H atom VUV LIF signals measured for the C + $H_2$, $C_3H_4$ and $CH_3OH$ reactions (in the presence of excess $H_2$, $C_3H_4$ and $CH_3OH$ respectively) at 127 K are shown in Figure 2. H atom traces were recorded in groups of three in successive experiments, repeated five times. The order in which traces were acquired was also varied to minimize errors arising from potential variations in the H atom signals as a function of acquisition time (mainly due to changes in the probe laser wavelength and a slow VUV energy drift due to impurities in the tripling cell). These traces are seen to contain two component parts; a rapid initial increasing part, due to the fast formation of atomic hydrogen by the reaction under investigation followed by a slow decaying part due to its diffusional loss from the probe volume in addition to any other secondary loss processes. The slow decaying part of each of these traces was fitted using a single exponential function which was then extrapolated to zero time to obtain an estimate of the nascent H atom signal amplitude from each individual trace. H atom signals recorded in the presence of excess $H_2$ trace the H atom yield of the $C(^1D) + H_2$ reaction alone, whereas the H atom signals recorded in the presence of $C_3H_4$ and $CH_3OH$ represent the H atom yield from the $C(^3P)/C(^1D) + C_3H_4$ and the $C(^3P)/C(^1D) + CH_3OH$ reactions respectively, therefore containing contributions from excited state carbon reactions.

Test experiments (described in the experimental section) showed that the $C(^1D) + C_3H_4$ reaction resulted in an atomic hydrogen yield of approximately 100%. Similar test measurements performed on the $C(^3P)/C(^1D) + CH_3OH$ system in the presence and absence of a large excess of $H_2$ also showed that the H atom signal did not change, indicating that the H atom yield of the $C(^1D) + CH_3OH$ reaction is also approximately 100% within the experimental uncertainties. It can be seen from Figure 2 that the long-time part of the H atom product trace for the $C(^3P)/C(^1D) + CH_3OH$ reaction (shown in red) decays more slowly than the corresponding H atom traces for the corresponding $C(^3P)/C(^1D) + C_3H_4$ and C($^1$D) + H$_2$ reactions. This effect can almost certainly be attributed to the slow decomposition of product HCO to H + CO during the timescale of the experiment. While this secondary H atom production leads to a slight overestimation of the primary H atom signal amplitude, it is somewhat offset by fitting to the slower long-time decay leading to a smaller H-atom signal at time zero. As a result, it is expected that this effect should not lead to significant errors in the H-atom yield from the C($^3$P) + CH$_3$OH reaction. The relative H atom yields from the C($^3$P) + CH$_3$OH reaction were obtained from the following expression

$$H_{C+CH_3OH} - H_{C(^1D)+H_2} \big/ ((H_{C+C_3H_4} - H_{C(^1D)+H_2})/0.85)$$

where $H_{C(^1D)+H_2}$, $H_{C+C_3H_4}$ and $H_{C+CH_3OH}$ are the atomic hydrogen signal amplitudes derived from fitting to the slow decaying part of H atom profiles recorded for C($^1$D)/C($^3$P) in the presence of H$_2$, C$_3$H$_4$ and CH$_3$OH respectively. After subtraction of the H atom signal originating from the C($^1$D) + C$_3$H$_4$ reaction, the H atom intensity recorded in the C + C$_3$H$_4$ experiments was corrected by a factor 0.85 to take into account the experimental H atom product yield for this process at room temperature.[34] A small correction factor was also applied to account for absorption of the VUV excitation and emission intensities by C$_3$H$_4$ which was calculated to be non-negligible. The mean value of the five corrected H atom yields for the C($^3$P) + CH$_3$OH reaction was then calculated, using the combined statistical uncertainties to give a weighted average value. At 296 K, the H atom yield of the C($^3$P) + CH$_3$OH reaction was seen to represent only (9.7 ± 2.7)% of the total. Similarly, at 127 K the H atom yield of the C($^3$P) + CH$_3$OH reaction was seen to represent only (10.2 ± 5.1)% of the total.

**MESMER calculations**

The experimental product branching ratio results described above allowed us to provide certain additional constraints for the initial parameters of the MESMER statistical calculations. Firstly, as the H atom yield was very low, it is clear that the reactive pathway involving addition of ground state atomic carbon to the carbon atom of methanol (passing through weakly bound complex2 and leading to H atom formation directly) is a very minor channel, contributing a maximum of 10-15% to the total reactivity. Secondly, as the H-atom yield did not vary significantly as the temperature was lowered to 127 K, there was little or no additional stabilization of complex2 at low temperature indicating that the branching fraction for this channel remains small and constant as a function of temperature. Indeed we would expect to see a marked increase in the H-atom yield if there was significant stabilization for complex2 at low temperature. If the reactive pathway passing through complex2 is a minor one, the channel involving addition of ground state atomic carbon to the oxygen atom of methanol (passing through complex1) must represent the major reaction pathway, yielding $CH_3$ and HCO which can fall apart further to give H + CO products. Using these constraints, the barrier heights of TS1 and TS4 were varied to fit the experimental data. Considering the H atom yield experiments, it was found that TS4 needed to be raised to at least 7 kJ mol$^{-1}$ above TS1 to predict a 15% yield down the red channel shown in Figure 1 at 296 K. This energy for TS4 represents a lower limit since the dissociation of HCO is a secondary source of H atoms as discussed above. Moreover as the H atom yield is observed to be approximately independent as a function of temperature, this result is inconsistent with any significant H atom production occurring via TS4 considering the necessarily larger barrier we have assigned to this channel. In summary the theoretical calculations suggest that the only significant channel in this reaction is HCO formation via TS1 and that despite the large exothermicity of this preferred pathway (the $CH_3$ + HCO products lie 326 kJ mol$^{-1}$ lower in energy than the reagent asymptote), 10-15% at most of the HCO radicals formed fall apart to give H + CO (assuming that all of the reactive flux passes through complex1). Given all these considerations, the channel via TS4 (shown in red in Figure 1) was considered to be very minor and was thus excluded from subsequent master equation calculations.

Considering only the black channel in Figure 1, the VRC calculations predict a modest negative temperature dependence of the capture rate with a value of $1.6 \times 10^{-10}$ cm$^3$ molecule$^{-1}$ s$^{-1}$ at 300 K increasing to 2.0 × $10^{-10}$ cm$^3$ molecule$^{-1}$ s$^{-1}$ at 50 K. It is noted that these VRC calculations have relatively large uncertainties, and when comparing initial theoretical results with the experimental data, it was found that the agreement between the two datasets was improved when the negative temperature dependence of the capture step was increased. In light of this effect, the ILT *A* factor was set to 1.6 × $10^{-10}$ cm$^3$ molecule$^{-1}$ s$^{-1}$ at 300 K whilst the $n^\infty$ parameter was allowed to vary between -0.5 and -0.1 in order to fit the experimental data. Given the large range of barrier heights obtained for TS1 at the different levels of theory used in this study, this value was also varied in order to give a "best fit" barrier height. Calculations were performed using both the M06-2X and CCSD(T)-F12 values for the energy of complex1.

When using the M06-2X well depth of -69 kJ mol$^{-1}$ the master equation calculations begin to predict substantial formation of stabilised complex1. In order to keep the amount of stabilised complex1 below 5%, consistent with the temperature dependence of the experimental H atom yields, the energy transfer parameter <$\Delta E_{down}$> was set to 35 cm$^{-1}$. For consistency, the same value was used for the calculations with the CCSD(T) well depth though it is noted that these calculations are relatively insensitive to changes in <$\Delta E_{down}$> and doubling the value of <$\Delta E_{down}$> to 70 cm$^{-1}$ resulted in only a 1% increase in the theoretical rate constant at 50 K.

**Kinetic results**

Examples of kinetic decays obtained by both chemiluminescence and VUV LIF detection methods at 296 K and 127 K are shown in Figure 3.

Methanol was held in large excess with respect to atomic carbon for all experiments so that simple exponential fits to the temporal profiles should have yielded pseudo-first-order rate constants ($k_{1st}$) for the C + CH$_3$OH reaction directly. Nevertheless, Figure 3 clearly shows that fits to NO* chemiluminescence decays were biexponential, containing a fast early time component representing a large fraction of the signal amplitude with a slower component at longer times. As C($^1$D) atoms are also formed in the flow by $CBr_4$ photolysis at the 10-20% level with respect to $C(^3P)$, the early time NO* signal originates from the $C(^1D) + NO_2 \rightarrow CO + NO$ reaction. Indeed, the NO* yield from the excited state reaction appears to overwhelm the NO* signal from the ground state one. Although the $C(^3P)$ and $NO_2(^2A_1)$ reactants correlate with $NO(B^2\Pi_r)$ and $CO(X^1\Sigma^+)$ products in $C_s$ symmetry, by far the most likely (exothermic) products are the ground state ones $NO(X^2\Pi_r)$ and $CO(X^1\Sigma^+)$ (544 kJ mol$^{-1}$ lower in energy than the $NO(B^2\Pi_r)$ and $CO(X^1\Sigma^+)$ products). In contrast, $C(^1D)$ and $NO_2(^2A_1)$ reactants correlate to the excited state $NO(A^2\Sigma^+ / B^2\Pi_r)$ and $CO(X^1\Sigma^+)$ products (the $B^2\Pi_r$ and $A^2\Sigma^+$ states of NO are separated by only 15 kJ mol$^{-1}$) so that we would expect a much larger fraction of $NO(B^2\Pi_r)$ state formation.[31] Consequently, the fast decaying part of the chemiluminescence trace thus followed the sum of $C(^1D)$ losses through the reactions of $C(^1D)$ with both $NO_2$ and $CH_3OH$ in addition to its losses by relaxation. As $NO_2$ was constant for any series of measurements and $C(^1D)$ relaxation was dominated by collisions with the carrier gas which was also constant, any change in the $k_{1st}$ value was thus due to reaction with methanol. At 296 K, the two exponential parts of the kinetic decay are essentially decoupled; the rate constant for the $C(^1D)$ + methanol reaction is clearly substantially larger than the $C(^3P)$ + methanol one allowing us to extract rate constants for both the ground and excited state carbon reactions. Below 127 K, however, the situation was quite different. Here, the two decays occurred on very similar timescales (incidentally indicating that the rates for the two reactions converge at lower temperature) so that it was no longer possible to decouple the two constituent processes. As a result, it was not possible to use the chemiluminescence detection method at temperatures lower than 127 K to extract rate constants. No attempt was made to extract the $C(^3P)$ + $CH_3OH$ reaction rate constant by chemiluminescence at 127 K as this involved analysis of the weak signal in the tail of the decay profiles. Moreover, little or no biexponential behavior was observed in chemiluminescence measurements recorded with the 177 K $N_2$ nozzle, indicating that $N_2$ acts as a more efficient quencher of $C(^1D)$ than argon. As a result it was not possible to follow $C(^1D)$ kinetics with nitrogen based flows. Conversely, kinetic decays could be obtained using the direct C($^3$P) VUV LIF detection method at all temperatures and with all carrier gases. In contrast to the chemiluminescence measurements these profiles are seen in Figure 3 to be well described by single exponential decays.

A range of methanol concentrations was used at each temperature (measurements were also performed in the absence of methanol) and the resulting $k_{1st}$ values were plotted against the corresponding methanol concentration. Examples of such plots are shown in Figure 4 with their slopes (weighted by the statistical uncertainties of individual $k_{1st}$ values) yielding rate constants for the C($^3$P)/C($^1$D) + CH$_3$OH reactions at specified temperatures.

The exploitable ranges of CH$_3$OH concentrations at 50 K and 75 K were severely limited by apparent fall-offs in the observed $k_{1st}$ values at high methanol concentrations. This effect was interpreted as the onset of cluster formation leading to a reduction in the free gas-phase CH$_3$OH concentrations, in a similar manner to our previous experiments on the reactivity of H$_2$O at low temperature.[29] Consequently, only experiments conducted in the linear regime were used in the final analysis.

Direct C($^3$P) VUV LIF detection allowed us to investigate the C($^3$P) + CH$_3$OH reaction over the range 50-296 K. Test measurements of the C($^3$P) + CH$_3$OH reaction at 127 K in the presence of a large excess of H$_2$ ([H$_2$] = 3.2 × 10$^{14}$ molecule cm$^{-3}$, (thereby removing C($^1$D) rapidly from the flow) yielded the same value for the second-order rate constant as the one obtained in the absence of H$_2$, indicating the negligible influence of C($^1$D) on C($^3$P) decays recorded by the VUV LIF method. Chemiluminescence methods allowed us to obtain rate constants for the C($^3$P) + CH$_3$OH reaction at 177 K and 296 K (thereby allowing a direct comparison of the two techniques). The chemiluminescence method yields rate constants for the C($^3$P) + CH$_3$OH reaction which are slightly lower than those derived from direct LIF measurements. This difference is almost certainly due to a contribution to the long-time part of the decays (which are used to extract C($^3$P) rate constants) from

NO* produced by the C($^1$D) + NO$_2$ reaction. As the yield of NO* seems to be significantly larger than the one from the corresponding C($^3$P) + NO$_2$ reaction, even a small amount of residual C($^1$D) atoms at long times could potentially interfere.

As the C($^1$D) + CH$_3$OH reaction occurred on a shorter timescale than the C($^3$P) + CH$_3$OH one (at 127 K and above), rate constants for the C($^1$D) + CH$_3$OH reaction could be measured at 127 K and 296 K using the chemiluminescence method. The measured values of around $1.7 \times 10^{-10}$ cm$^3$ molecule$^{-1}$ s$^{-1}$ indicate that this reaction exhibits little or no temperature dependence, in a similar manner to the C($^1$D) + H$_2$ reaction which has been studied both experimentally at 300 K[33] and theoretically[50] down to low temperatures, with rate constants in the range (1-2) × 10$^{-10}$ cm$^3$ molecule$^{-1}$ s$^{-1}$ above 100 K.

Measurements of the kinetics of the C($^3$P) + CH$_3$OH reaction were also performed at 241 K using a Laval nozzle based on a mixture of SF$_6$ and N$_2$. Our previous work on the CH + H$_2$O reaction[29] has demonstrated that the use of SF$_6$ in particular as the carrier gas can result in an enhancement in the measured rate at low temperature, compared to experiments employing Argon, almost certainly due to the efficient collisional stabilization of complex1. It can be seen from Figure 4 that a small enhancement of the rate constant (a 50% increase of the rate constant with respect to the calculated trend for Argon) is observed in this case although the magnitude of the effect is lower than the corresponding effect for the CH + H$_2$O reaction at the same temperature.[29]

In order to explore the difference between the experimental results using Ar and SF$_6$ carrier gases, calculations were performed using SF$_6$ as a bath gas. Given the relative sizes of the two bath gases, collision energy transfer is expected to be more efficient for SF$_6$ compared to Ar. Calculations were performed at 241 K using a well depth for complex1 of -69 kJ mol$^{-1}$. In order to predict the observed 50% increase, it was necessary to increase the <$\Delta E_{down}$> value to 500 cm$^{-1}$.

The temperature dependent rate constants obtained for both the ground and first excited state reactions and by both detection methods are presented in Figure 5 and are summarized in Table 2 alongside the results of the MESMER calculated rates for the C($^3$P) + CH$_3$OH reaction. It was found that good agreement between experiment and theory could be obtained using both the M06 and CCSD(T) well depths for complex1 and these two models gave fitted energies for TS1 of -4.13 kJ mol$^{-1}$ and -4.12 kJ mol$^{-1}$ with n factors of -0.43 and -0.60 respectively. The significant negative temperature dependence observed for the rate constants of the C($^3$P) + CH$_3$OH reaction is apparent, and theoretically it is found that this can only be reproduced, through either significant stabilisation of complex1 or through a negative temperature dependence for the flux through the outer transition state. As the experimental H atom yields demonstrate that stabilisation into complex1 is relatively minor, a significant contribution from the outer transition state is crucial to obtain agreement between experiment and theory.

## 4. Conclusions and Astrophysical Implications

An experimental investigation of the C($^3$P) + CH$_3$OH reaction by both direct VUV LIF and indirect chemiluminescence detection methods has shown that the rate constant for this process increases as the temperature falls, becoming rapid at low temperature. The major products are shown to be HCO + CH$_3$, with a small fraction of the HCO radicals decomposing further to H + CO under the present experimental conditions through measurements of the relative H atom product yields. Statistical calculations of the low temperature rate constants are in excellent agreement with the experimental values after certain product channels are excluded by taking into consideration the experimental H atom branching ratios. Secondary measurements of the C($^1$D) + CH$_3$OH reaction using a chemiluminescence detection method have shown that the rate constant for this reaction is approximately independent of temperature over the 127 – 296 K range.

In order to test the impact of the C($^3$P) + CH$_3$OH reaction on gas-phase interstellar methanol abundances (the reaction is not present in current astrochemical models), the statistical calculations have been used to extrapolate the rate constant to temperatures and pressures more representative of those found in dense interstellar clouds. As it was not possible to perform these calculations at 10 K due to numerical difficulties caused by energy transfer transition probabilities becoming vanishingly small, the value obtained at 20 K was used instead. At 20 K and at the low pressure limit, the calculations predict that the rate constant should be close to the capture rate limiting value of $4.0 \times 10^{-10}$ cm$^3$ molecule$^{-1}$ s$^{-1}$. Although this may slightly underestimate the rate at 10 K, it is a reasonable approximation given the inherent uncertainty in the statistical calculations at such low temperatures. The C($^3$P) + CH$_3$OH reaction was introduced into the chemical network used by the astrochemical model Nautilus[51] leading to HCO and CH$_3$ as the exclusive products. Nautilus computes chemical abundances in the gas-phase and on the surface of interstellar grains as a function of time under the extreme conditions of the interstellar medium. In this particular case, the gas-grain interactions which mimic the uptake of gas-phase species are essential to simulate the presence of molecules such as methanol which are primarily formed on the surface of interstellar dust particles. Whilst the reactions of molecules on the surface of dust grains allows us to model the formation of larger more complex molecules on the interstellar grains themselves, they do not allow us to explain the presence of methanol in the gas-phase. Indeed, desorption mechanisms are required to return species formed in this way back to the gas-phase. Three processes are typically included; thermal desorption which obeys a standard Boltzmann law to overcome the energy barrier for the physisorbed molecules and cosmic ray desorption which involves whole grain heating through the collision of cosmic rays with dust grains. Both of these effects are too weak to be able to reproduce typical gas phase methanol abundances in dense clouds. The third desorption mechanism, namely non-thermal desorption, relies on the surface reaction energy to desorb the newly formed molecules. Models employing this reactive desorption approach assume a probability factor *f* for desorption which takes into consideration the exothermic energy of the reaction, the binding energy of the product molecule with the surface and the size of the newly formed species. Values of *f* of a few percent are generally required to reproduce gas phase methanol concentrations.[52] A fourth mechanism, namely photodesorption by UV photons produced deep in the dark cloud itself by the radiative decay of electronically excited $H_2$ could also be a non-negligible way to return molecules to the gas-phase[53] although these processes are not included in the current model.

Supplementary details on the processes included in the present model can be found in Semenov *et al.*[51] and Hersant *et al.*[54] The kinetic data and other parameters (physical conditions, elemental abundances and initial chemical composition) used for the simulations are the same as in those used in Loison *et al.*[55] for typical dark interstellar clouds. Two values of the oxygen elemental abundance have been employed ($2.4 \times 10^{-4}$ and $1.4 \times 10^{-4}$ with respect to total hydrogen) resulting in C/O elemental ratios of 0.7 and 1.2. The temperature is held at 10 K and the total density is $2 \times 10^4$ cm$^{-3}$. Figure 6 shows the relative abundances of $CH_3OH$ as a function of time obtained using the gas-grain model with the $C(^3P)$ + $CH_3OH$ reaction included compared to those obtained without this reaction present for the two nominal C/O ratios. For both C/O elemental ratios, the $CH_3OH$ abundance in the gas-phase is much smaller at early times (before $3 \times 10^5$ yr) as the high abundance of atomic carbon in the gas-phase destroys $CH_3OH$ extremely efficiently. After $3 \times 10^5$ yr, the $CH_3OH$ abundance is still slightly decreased but it is somewhat closer to previous predictions. Using the reactive flux (equal to the product of the rate constant and the concentrations of carbon and methanol respectively) as the criterion for classifying the importance of the various gas-phase methanol destruction mechanisms, our simulations indicate that the $C(^3P)$ + $CH_3OH$ reaction is the most efficient gas-phase methanol destruction mechanism between $1 \times 10^4$ yr and $3 \times 10^6$ yr for both C/O ratios and can be as much as three orders of magnitude greater than the next most significant methanol loss process. The abundances of $CH_3$ and HCO are relatively unchanged by the introduction of the $C(^3P)$ + $CH_3OH$ reaction as these species have many sources in current astrochemical models.

This result has important consequences for the gas-grain chemistry of current astrochemical models. The presence of large abundances of atomic carbon in the gas-phase in dark clouds (see for example Schilke *et al.*[21]) could reduce strongly the abundance of methanol. Observations of large abundances of methanol in the gas-phase (of a few $10^{-9}$ compared to the total hydrogen density) would then be an indicator of dark clouds older than a few $10^5$ yr, for which the atomic carbon has had time to be converted into other C-bearing species. Gas phase methanol abundances in the range $(1-3) \times 10^{-9}$ have been reported for dense clouds TMC-1[56] and L134N.[57] Since the abundance at the surface of the grains is larger by orders of magnitude, another explanation could be that methanol desorption mechanisms are more efficient than currently estimated. Recent experiments[58] suggest that reactive desorption mechanisms could return larger percentages of species formed on ice grains to the gas-phase than the values used in current models although this effect depends to a large extent on the composition and morphology of the surface itself.

Alternatively, there could be reactive pathways other than the hydrogenation of CO, which might lead to methanol formation on interstellar ices. One potential source of methanol could begin with the barrierless reaction of C with H on the grain surface leading to CH radical formation. If the CH radicals are formed in sites adjacent to water molecules already present on the surface, it may be possible that reaction between these two species could occur efficiently. Indeed, experiments[29] and calculations[59] of the gas-phase reactivity of CH with $H_2O$ indicate that this reaction proceeds in the absence of a barrier and is fast at low temperatures, also leading to the formation of $H_2CO$. As the initial hydrogenation step of the standard methanol formation mechanism (H + CO to form HCO) is characterized by an activation barrier of approximately 19 kJ mol$^{-1}$,[60] and is therefore rate limiting, this secondary source of $H_2CO$ could then lead to the formation of significant quantities of methanol through the subsequent hydrogenation of $H_2CO$.

**Acknowledgements**

JCL, PC and KMH were supported by the Conseil Régional d'Aquitaine (20091102002), the INSU-CNRS national programs PCMI and PNP, the Observatoire Aquitain des Sciences de l'Univers and the Hubert Curien Program Galileo project number 28125ZM. NB thanks the bilateral project UIF/UFI - Galileo Project 2013 (G12-70) between Italy and France for supporting her stay in Bordeaux. The astrochemical modelling work was supported by an ERC Starting Grant (3DICE, grant agreement 336474).

[a]*School of Chemistry, University of Leeds, Leeds, LS2 9JT, UK*

[b]*Université de Bordeaux, Laboratoire d'Astrophysique de Bordeaux, UMR 5804, F-33270 Floirac, France.*

[c]*CNRS, Laboratoire d'Astrophysique de Bordeaux, UMR 5804, F-33270 Floirac, France.*

[d]*Université de Bordeaux, Institut des Sciences Moléculaires, UMR 5255, F-33400 Talence, France. Tel: +33 5 40 00 63 42.*

[e]*CNRS, Institut des Sciences Moléculaires, UMR 5255, F-33400 Talence, France.*

[f]*Dipartimento di Chimica, Biologia e Biotecnologie, Universita' degli Studi di Perugia, 06123 Perugia, Italy*

[g]*National Centre for Atmospheric Science, University of Leeds, Leeds, LS2 9JT, UK*

E-mail: [km.hickson@ism.u-bordeaux1.fr](km.hickson@ism.u-bordeaux1.fr)

## References

1  D. C. Clary, N. Haider, D. Husain and M. Kabir, *Astrophys. J.* 1994, **422**, 416.

2  D. Chastaing, P. L. James, I. R. Sims and I. W. M. Smith, *Phys. Chem. Chem. Phys.*, 1999, **1**, 2247.

3  D. Chastaing, S. D. Le Picard and I. R. Sims, *J. Chem. Phys.*, 2000, **112**, 8466.

4  D. Chastaing, S. D. Le Picard, I. R. Sims and I. W. M. Smith, *Astron. Astrophys.*, 2001, **365**, 241.

5   D. C. Clary, E. Buonomo, I. R. Sims, I. W. M. Smith, W. D. Geppert, C. Naulin, M. Costes, L. Cartechini and P. Casavecchia, *J. Phys. Chem. A*, 2002, **106**, 5541.

6   C. Ochsenfeld, R. I. Kaiser, Y. T. Lee, A. G. Suits and M. Head-Gordon, *J. Chem. Phys.* 1997, **106**, 4141.

7   R. I. Kaiser and A. M. Mebel, *Int. Rev. Phys. Chem.*, 2002, **21**, 307.

8   F. Leonori, R. Petrucci, E. Segoloni, A. Bergeat, K. M. Hickson, N. Balucani and P. Casavecchia, *J. Phys. Chem. A*, 2008, **112**, 1363.

9   M. Costes, P. Halvick, K. M. Hickson, N. Daugey and C. Naulin, *Astrophys. J.* 2009, **703**, 1179.

10  C. Naulin, N. Daugey, K. M. Hickson and M. Costes, *J. Phys. Chem. A*, 2009, **113**, 14447.

11  E. S. Wirström, W. D. Geppert, Å. Hjalmarson, C. M. Persson, J. H. Black, P. Bergman, T. J. Millar, M. Hamberg and E. Vigren, *Astron. Astrophys.*, 2011, **533**, A24.

12  B. E. Turner, *Astrophys. J.* 1998, **501**, 731.

13  P. Friberg, S. C. Madden, Å. Hjalmarson and W. M. Irvine, *Astron. Astrophys.*, 1988, **195**, 281.

14  W. D. Geppert, M. Hamberg, R. D. Thomas, F. Österdahl, F. Hellberg, V. Zhaunerchyk, A. Ehlerding, T. J. Millar, H. Roberts, J. Semaniak, M. af Ugglas, A. Källberg, A. Simonsson, M. Kaminska and M. Larsson, *Faraday Discuss.*, 2006, **133**, 177.

15  S. Bottinelli, A. C. Adwin Boogert, J. Bouwman, M. Beckwith, E. F. van Dishoeck, K. I. Öberg, K. M. Pontoppidan, H. Linnartz, G. A. Blake, N. J. Evans II and F. Lahuis, *Astrophys. J.* 2010, **718**, 1100.

16  N. Watanabe, A. Nagaoka, T. Shiraki and A. Kouchi, 2004, *Astrophys. J.* 2004, **616**, 638.

17  G. W. Fuchs, H. M. Cuppen, S. Ioppolo, C. Romanzin, S. E. Bisschop, S. Andersson, E. F. van Dishoeck and H. Linnartz, *Astron. Astrophys.*, 2009, **505**, 629.

18  R. Garrod, I. H. Park, P. Caselli and E. Herbst, *Faraday Discuss.*, 2006, **133**, 51.

19  A. I. Vasyunin and E. Herbst, *Astrophys. J.* 2013, **769**, 34.

20  T. G. Phillips and P. J. Huggins, *Astrophys. J.* 1981, **251**, 533

21 P. Schilke, J. Keene, J. Le Bourlot, G. Pineau des Forêts and E. Roueff, *Astron. Astrophys.*, 1995, **294**, L17.

22 J. Daranlot, U. Hincelin, A. Bergeat, M. Costes, J.-C. Loison, V. Wakelam and K. M. Hickson, *Proc. Natl. Acad. Sci. USA*, 2012, **109**, 10233.

23 Y. Dede and I. Ozkan, *Phys. Chem. Chem. Phys.*, 2012, **14**, 2326.

24 D. R. Glowacki, C.-H. Liang, C. Morley, M. J. Pilling and S. H. Robertson, *J. Phys. Chem. A*, 2012, **116**, 9545.

25 R. J. Shannon, M. A. Blitz, A. Goddard and D. E. Heard, *Nature Chem.*, 2013, **5**, 745.

26 R. J. Shannon, R. L. Caravan, M. A. Blitz and D. E. Heard, *Phys. Chem. Chem. Phys.*, 2014, **16**, 3466.

27 B. R. Rowe, G. Dupeyrat, J. B. Marquette and P. Gaucherel, *J. Chem. Phys.*, 1984, **80**, 4915.

28 N. Daugey, P. Caubet, A. Bergeat, M. Costes and K. M. Hickson, *Phys. Chem. Chem. Phys.,* 2008, **10**, 729.

29 K. M. Hickson, P. Caubet and J.-C. Loison, *J. Phys. Chem. Lett*. 2013, **4**, 2843.

30 B.-M. Cheng, M. Bahou, Y.-P. Lee and L. C. Lee, *J. Chem. Phys.*, 2002, **117**, 1633.

31 G. Dorthe, J. Caille, S. Burdenski, P. Caubet, M. Costes and G. Nouchi, *J. Chem. Phys.*, 1985, **82**, 2313.

32 A. Kramida, Yu. Ralchenko, J. Reader and NIST ASD Team, NIST Atomic Spectra Database, version 5.1, 2013.National Institute of Standards and Technology, Gaithersburg, MD. http://physics.nist.gov/asd

33 K. Sato, N. Ishida, T. Kurakata, A. Iwasaki and S. Tsunashima, *Chem. Phys.*, 1998, **237**, 195.

34 J.-C. Loison and A. Bergeat, *Phys. Chem. Chem. Phys.,* 2004, **6**, 5396.

35 Y. Zhao and D. G. Truhlar, *Theor. Chem. Acc.*, 2008, **120**, 215.

36 G. -S. Kim, T. L. Nguyen, A. M. Mebel, S. H. Lin and M. T. Nguyen, *J. Phys. Chem. A*, 2003, **107**, 1788.

37 P. R. Schreiner and H. P. Reisenauer, *ChemPhysChem*, 2006, **7**, 880.

38 A. F. Wagner, *Proceedings of the Combustion Institute*, 2002, **29**, 1173.

39 D. L. Baulch, C. T. Bowman, C. J. Cobos, R. A. Cox, Th. Just, J. A. Kerr, M. J. Pilling, D. Stocker, J. Troe, W. Tsang, R. W. Walker and J. Warnatz, *J. Phys. Chem. Ref. Data*, 2005, **34**, 757.

40 I. Ozkan and Y. Dede, *Int. J. Quantum Chem.*, 2012, **112**, 1165.

41 T. J. Lee, *Chem. Phys. Lett.*, 2003, **372**, 362.

42 M. L. Leininger, I. M. B. Nielsen, T. D. Crawford and C. L. Janssen, *Chem. Phys. Lett.*, 2000, **328**, 431.

43 S. A. Carr, D. R. Glowacki, C.-H. Liang, M. T. Baeza Romero, M. A. Blitz, M. J. Pilling and P. W. Seakins, *J. Phys. Chem. A*, 2011, **115**, 1069.

44 M. A. Blitz, K. J. Hughes, M. J. Pilling and S. H. Robertson, *J. Phys. Chem. A*, 2006, **110**, 2996.

45 J. W. Davies, N. J. B. Green and M. J. Pilling, *Chem. Phys. Lett.*, 1986, **126**, 373.

46 Y. Georgievskii and S. J. Klippenstein, *J. Chem. Phys.*, 2003, **118**, 5442.

47 Y. Georgievskii and S. J. Klippenstein, *J. Phys. Chem. A*, 2003, **107**, 9776.

48 J. Zheng, S. Zhang, B. J. Lynch, J. C. Corchado, Y.-Y. Chuang, P. L. Fast, W.-P. Hu, Y.-P. Liu, G. C. Lynch, K. A. Nguyen, C. F. Jackels, A. Fernandez Ramos, B. A. Ellingson, V. S. Melissas, J. Villa, I. Rossi, E. L. Coitino, J. Pu, T. V. Albu, R. Steckler, B. C. Garrett, A. D. Isaacson and D. G. Truhlar, POLYRATE-version 2008, University of Minnesota, Minneapolis, 2008. http://comp.chem.umn.edu/polyrate/

49 J. Zheng, S. Zhang, J. C. Corchado, Y.-Y. Chuang, E. L. Coitino, B. A. Ellingson, and D. G. Truhlar, GAUSSRATE version 2009-A/P2008-G09/G03/G98/G94, University of Minnesota, Minneapolis, 2009. http://t1.chem.umn.edu/gaussrate/

50 S. Y. Lin and H. Guo, *J. Phys. Chem. A*, 2004, **108**, 10066.

51 D. Semenov, F. Hersant, V. Wakelam, A. Dutrey, E. Chapillon, S. Guilloteau, T. Henning, R. Launhardt, V. Piétu and K. Schreyer, *Astron. Astrophys.*, 2010, **522**, A42.

52 R. T. Garrod, V. Wakelam and E. Herbst, *Astron. Astrophys.*, 2007, **467**, 1103.

53 M. Bertin, E. C. Fayolle, C. Romanzin, H. A. M. Poderoso, X. Michaut, L. Philippe, P. Jeseck, K. I. Oberg, H. Linnartz and J.-H. Fillion, *Astrophys. J.*, 2013, **779**, 120.

54 F. Hersant, V. Wakelam, A. Dutrey, S. Guilloteau and E. Herbst, *Astron. Astrophys.*, 2009, **493**, L49.

55 J.-C. Loison, V. Wakelam, K. M. Hickson, A. Bergeat and R. Mereau, *Mon. Not. Royal Astron. Soc.*, 2014, **437**, 930.

56 P. Pratap, J. E. Dickens, R. L. Snell, M. P. Miralles, E. A. Bergin, W. M. Irvine and F. P. Schloerb, *Astrophys. J.*, 1997, **486**, 862.

57 J. E. Dickens, W. M. Irvine, R. L. Snell, E. A. Bergin, F. P. Schloerb, P. Pratap and M. P. Miralles, *Astrophys. J.*, 2000, **542**, 870.

58 F. Dulieu, E. Congiu, J. Noble, S. Baouche, H. Chaabouni, A. Moudens, M. Minissale and S. Cazaux, *Sci. Rep.,* 2013, **3**, 1338.

59 A. Bergeat, S. Moisan, R. Mereau and J.-C. Loison, *Chem. Phys. Lett.*, 2009, **480**, 21.

60 P. S. Peters, D. Duflot, L. Wiesenfeld and C. Toubin, *J. Chem. Phys.*, 2013, **139**, 164310.

**Table 1** Continuous supersonic flow characteristics

| Laval nozzle Mach number | Mach2 (N$_2$/SF$_6$) 2.1 ± 0.01[†] | Mach2 N$_2$ 1.8 ± 0.02 | Mach2 N$_2$ (N$_2$/Ar) 1.9 ± 0.02 | Mach2 N$_2$ (using Ar) 2.0 ± 0.03 | Mach3 Ar 3.0 ± 0.1 | Mach4 Ar 3.9 ± 0.1 |
|---|---|---|---|---|---|---|
| Carrier gas | (N$_2$/SF$_6$)[‡] | N$_2$ | (N$_2$/Ar) | Ar | Ar | Ar |
| Density (× 10$^{16}$ cm$^{-3}$) | 9.0 | 9.4 | 10.4 | 12.6 | 14.7 | 25.9 |
| Impact pressure (Torr) | 11.5 | 8.2 | 9.1 | 10.5 | 15.3 | 29.6 |
| Stagnation pressure (Torr) | 18.6 | 10.3 | 12.0 | 13.9 | 34.9 | 113 |
| Temperature (K) | 241 ± 1 | 177 ± 2 | 158 ± 2 | 127 ± 2 | 75 ± 2 | 50 ± 1 |
| Mean flow velocity (ms$^{-1}$) | 282 ± 2 | 496 ± 4 | 456 ± 4 | 419 ± 3 | 479 ± 3 | 505 ± 1 |

[†] The errors on the Mach number, temperature and mean flow velocity, cited at the level of one standard deviation from the mean are calculated from separate measurements of the impact pressure using a Pitot tube as a function of distance from the Laval nozzle and the stagnation pressure within the reservoir.

[‡] Mole fraction N$_2$ / SF$_6$ = 0.4 / 0.6.

**Table 2** Measured and calculated rate constants for the C($^3$P)/C($^1$D) + CH$_3$OH reactions

| $T$ / K | [CH$_3$OH]$^\dagger$ | $N^\ddagger$ | Measured $k$ C($^3$P) + CH$_3$OH | Calculated $k$ C($^3$P) + CH$_3$OH | $N^\ddagger$ | Measured $k$ C($^1$D) + CH$_3$OH |
|---|---|---|---|---|---|---|
| 50 ± 1 | 0 - 4.3 | 36 | (20.0 ± 2.1)$^\S$ | 22$^\P$ | | |
| 75 ± 2 | 0 - 6.2 | 30 | (13.8 ± 1.5) | 13 | | |
| 127 ± 2 | 0 - 31 | 73 | (9.4 ± 1.0) | 6.9 | 39 | (16.8 ± 1.7)$^|$ |
| 158 ± 2 | 0 - 24 | 23 | (6.2 ± 0.7) | 5.3 | | |
| 177 ± 2 | 0 – 30 | 55 | (3.9 ± 0.4) | 4.6 | | |
| | 0 - 22 | 54 | (3.0 ± 0.3) | | | |
| 241 ± 1 | | 27 | (4.3 ± 0.5) | | | |
| 296 | 0 - 66 | 49 | (2.2 ± 0.3) | 2.9 | 39 | (16.3 ± 1.7) |
| | 0 - 36 | 41 | (1.6 ± 0.2) | | | |

$^\dagger$ / 10$^{13}$ molecule cm$^{-3}$.

$^\ddagger$Number of individual measurements.

$^\S$(20.0 ± 2.1) ≡ (20.0 ± 2.1) × 10$^{-11}$ cm$^3$ molecule$^{-1}$ s$^{-1}$.

$^\P$22 ≡ 22 × 10$^{-11}$ cm$^3$ molecule$^{-1}$ s$^{-1}$.

$^|$(16.8 ± 1.7) ≡ (16.8 ± 1.7) × 10$^{-11}$ cm$^3$ molecule$^{-1}$ s$^{-1}$.

Uncertainties on the measured rate constants represent the combined statistical (1σ) and estimated systematic (10%) errors. Uncertainties on the calculated temperatures represent the statistical (1σ) errors obtained from Pitot tube measurements of the impact pressure.

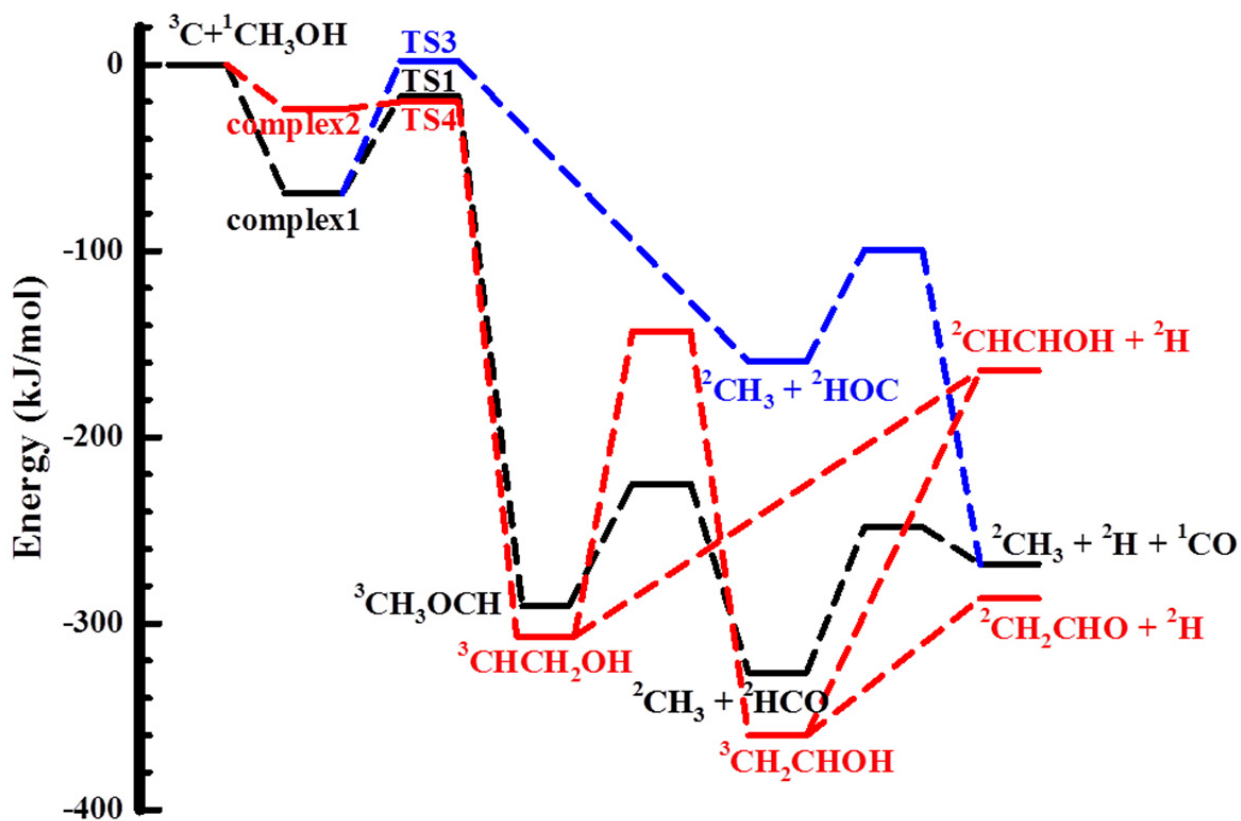

**Fig. 1**. Potential energy diagram for the C($^3$P) + methanol reaction on the triplet surface calculated at the M06-2X/cc-pVQZ level. See Table S1 for the relative energies. The TS1, TS3 and TS4 energies were adjusted during the Master Equation calculations (for details see text).

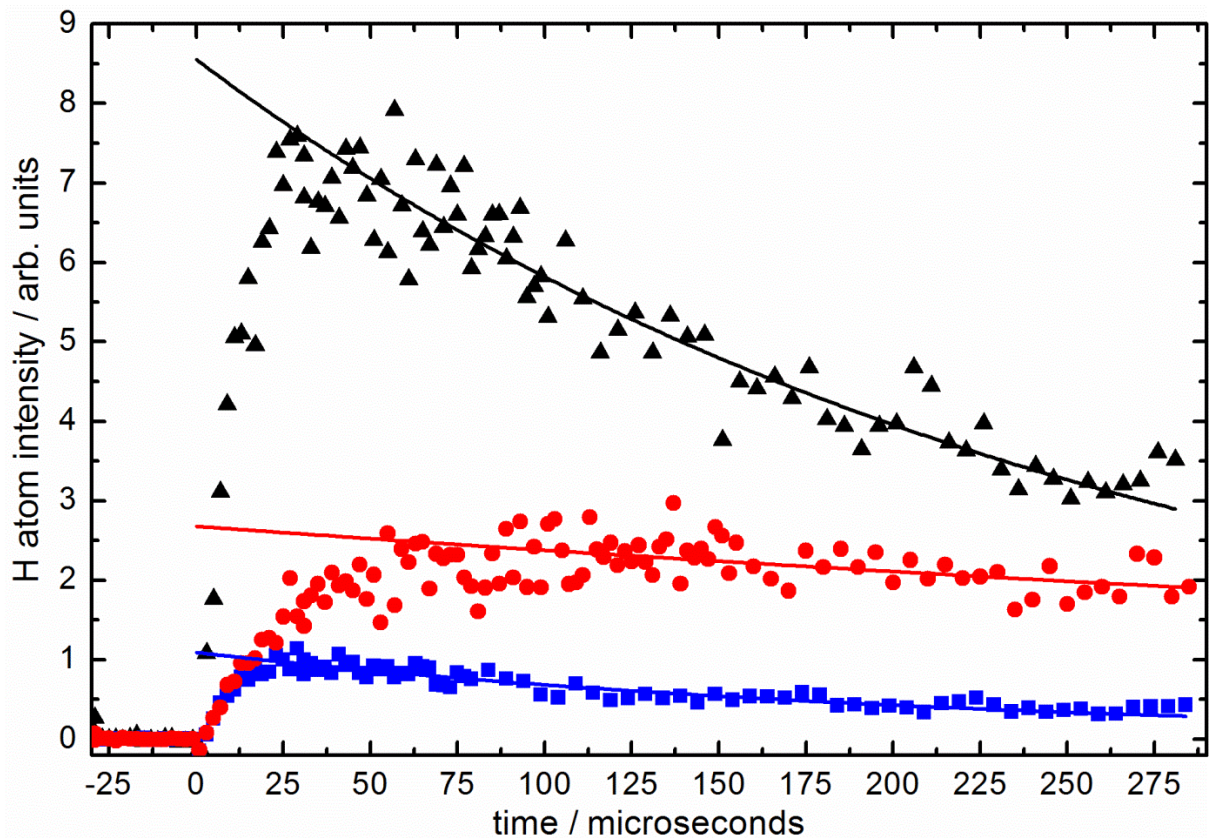

**Fig. 2**. Variation of the VUV LIF emission from H($^2$S) atoms produced by three different reactions at 127 K. (Black filled triangles) the C($^3$P)/($^1$D) + C$_3$H$_4$ reaction with [C$_3$H$_4$] = 2.9 × 10$^{14}$ molecule cm$^{-3}$; (Red filled circles) the C($^3$P)/($^1$D) + CH$_3$OH reaction with [CH$_3$OH] = 1.4 × 10$^{14}$ molecule cm$^{-3}$; (Blue filled squares) the C($^1$D) + H$_2$ reaction with [H$_2$] = 3.2 × 10$^{14}$ molecule cm$^{-3}$. The fits to the long-time part of each trace (representing the diffusional loss of H($^2$S)) were extrapolated to time zero to obtain the nascent H atom yields.

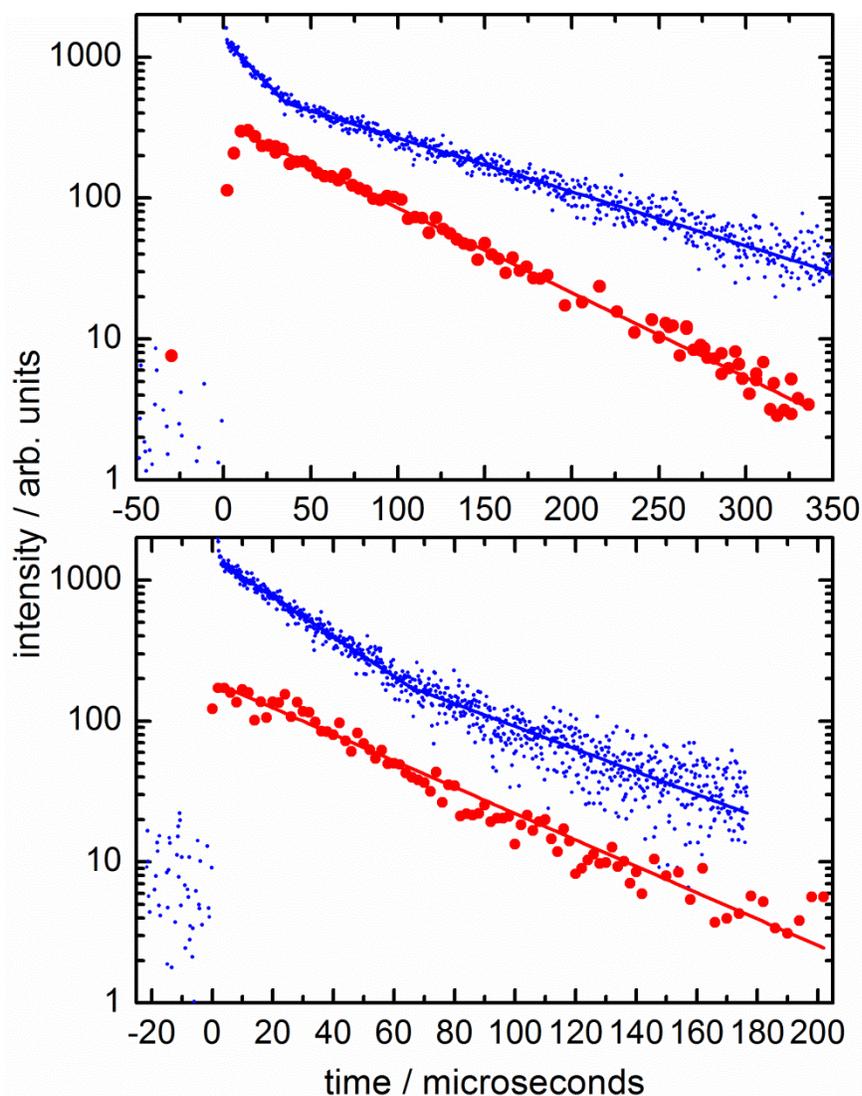

**Fig.3**. Variation of the chemiluminescence (blue filled circles) and VUV LIF (red filled circles) emission signals from reactant C($^3$P) atoms (NO* for the chemiluminescence measurements produced by the C + NO$_2$ reaction) as a function of time. (Upper panel) 296 K with [CH$_3$OH] = 1.2 × 10$^{14}$ molecule cm$^{-3}$; (lower panel) 127 K with [CH$_3$OH] = 2.0 × 10$^{14}$ molecule cm$^{-3}$. Biexponential behavior is clearly observed in chemiluminescence traces due to the fast C($^1$D) + CH$_3$OH reaction.

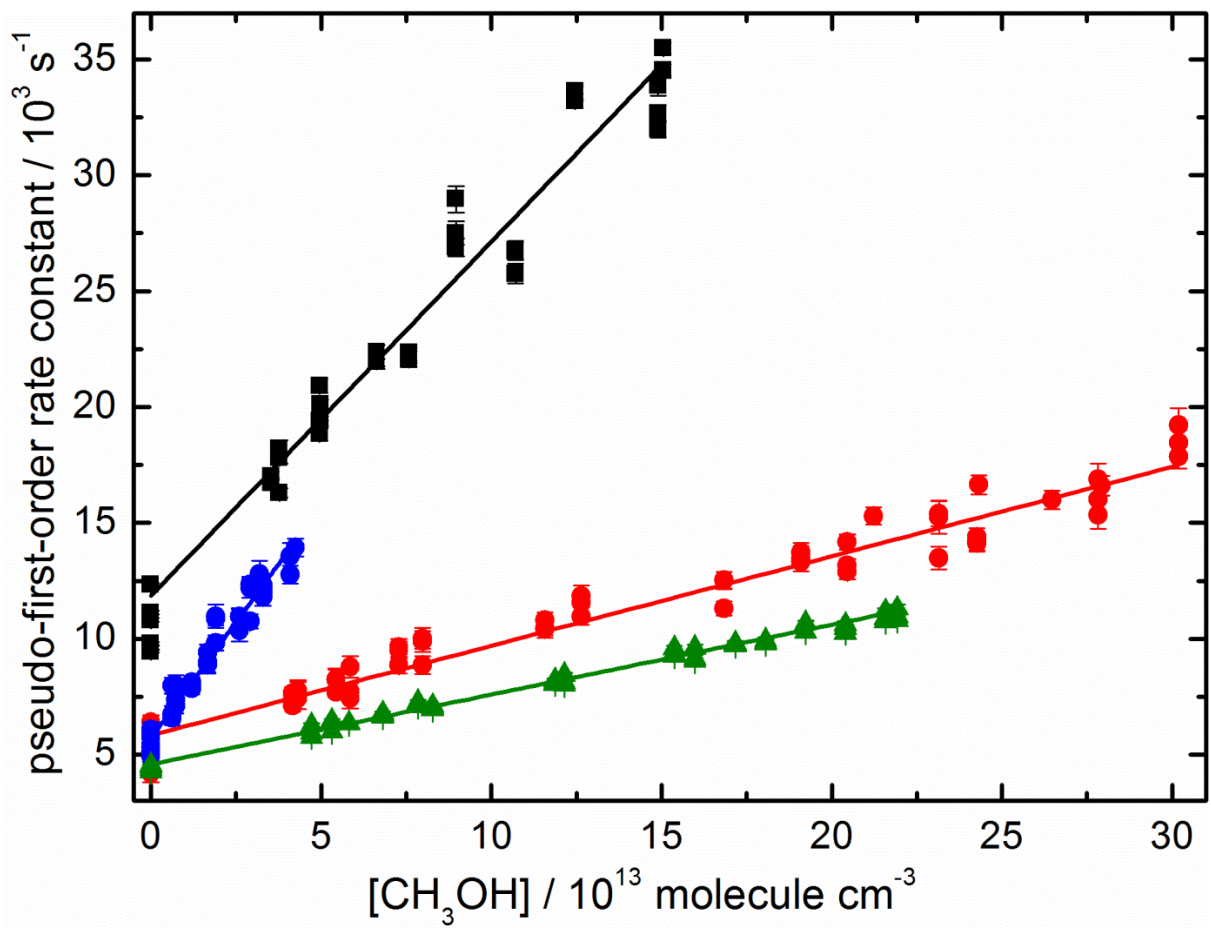

**Fig. 4**. Pseudo-first-order rate constants for the C($^3$P)/($^1$D) + CH$_3$OH reactions as a function of the methanol concentration. A weighted linear least squares fit yields the second-order rate constant. (Red filled circles) VUV LIF measurements of the rate constant for the C($^3$P) + CH$_3$OH reaction at 177 K; (green filled triangles) chemiluminescence measurements of the rate constant for the C($^3$P) + CH$_3$OH reaction at 177 K; (blue filled circles) VUV LIF measurements of the rate constant for the C($^3$P) + CH$_3$OH reaction at 50 K; (black filled squares) chemiluminescence measurements of the rate constant for the C($^1$D) + CH$_3$OH reaction at 127 K. The error bars reflect the statistical uncertainties at the level of a single standard deviation obtained by fitting to temporal profiles such as those shown in Fig. 3.

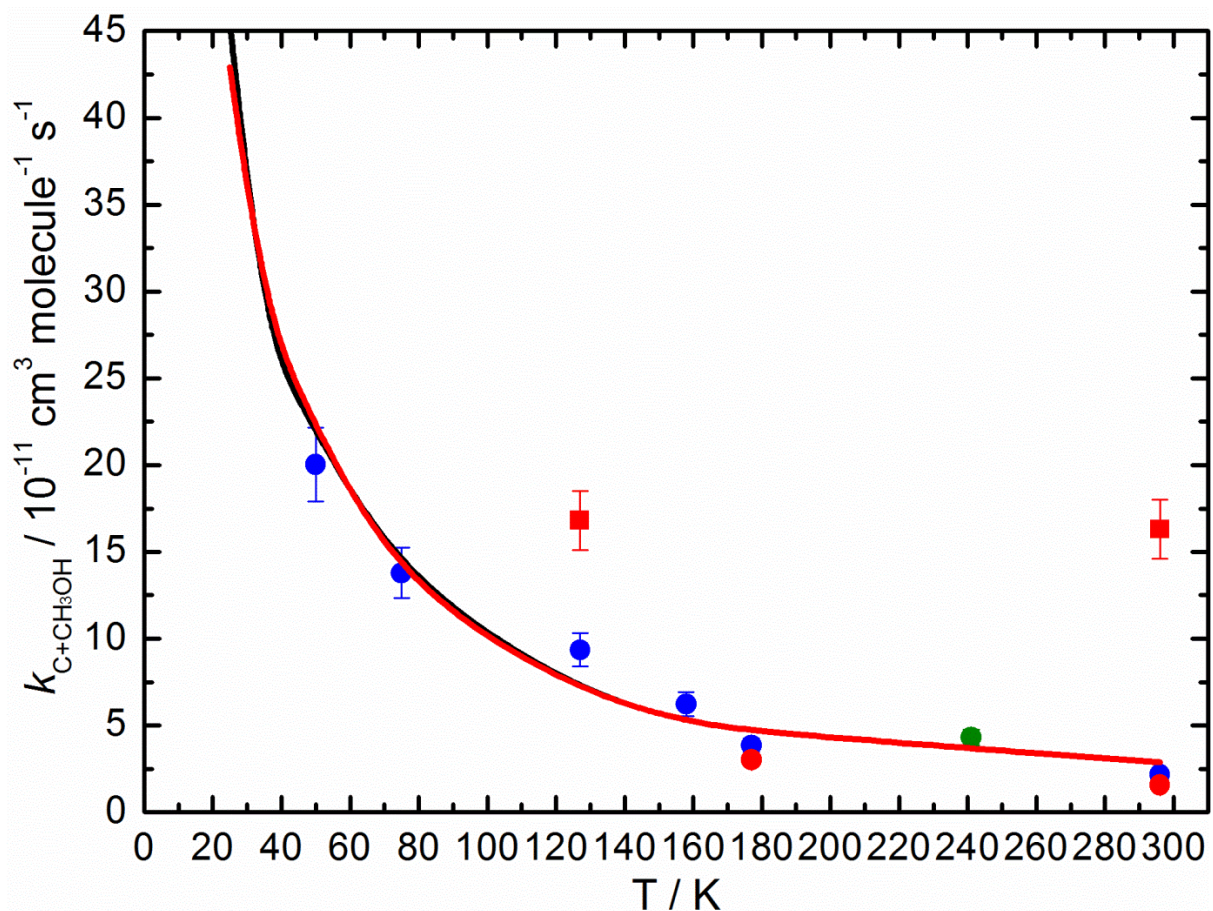

**Fig. 5**. Rate constants for the C($^3$P)/($^1$D) + CH$_3$OH reactions as a function of temperature. (Blue filled circles) VUV LIF measurements of the C($^3$P) + CH$_3$OH reaction with Ar or N$_2$ as the carrier gases; (green filled circles) VUV LIF measurements of the C($^3$P) + CH$_3$OH reaction with SF$_6$ / N$_2$ as the carrier gas mixture; (red filled circles) chemiluminescence measurements of the C($^3$P) + CH$_3$OH reaction; (red filled squares) chemiluminescence measurements of the C($^1$D) + CH$_3$OH reaction; (red solid line) MESMER statistical calculations of the rate constant for the C($^3$P) + CH$_3$OH reaction using a well depth for complex1 of -69 kJ mol$^{-1}$; (black solid line) MESMER statistical calculations using a well depth of -49 kJ mol$^{-1}$ for complex1. The error bars reflect the combined statistical and systematic (estimated at 10%) uncertainties.

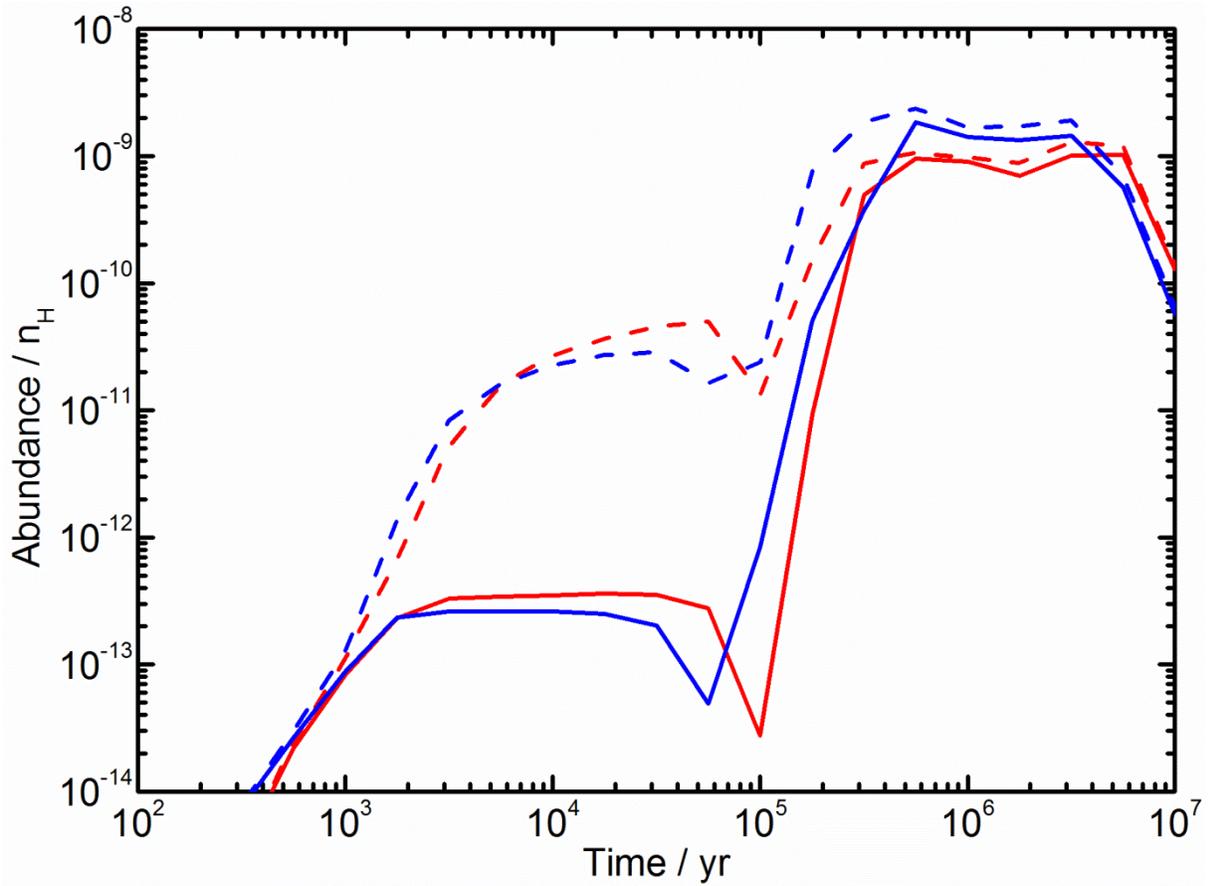

**Fig. 6**. Relative abundance of CH$_3$OH as a function of time using a dense cloud model. (Red lines) simulations using a value for C/O = 0.7. (Blue lines) simulations using a value for C/O = 1.2. (Dashed lines) simulations in the absence of the C+ CH$_3$OH reaction; (Solid lines) simulations using a value for the rate constant $k_{\text{C+CH3OH}}$ (10 K) = 4.0 × 10$^{-10}$ cm$^3$ molecule$^{-1}$ s$^{-1}$.